\documentclass[prb, twocolumn, superscriptaddress, showpacs,floatfix]{revtex4}  
\usepackage{hyperref}
\usepackage{amssymb}
\usepackage{amsmath}
\usepackage{float}
\usepackage{bbm}
\usepackage{graphicx}
\usepackage{epsfig}
\usepackage{epstopdf}
\usepackage{verbatim}
\usepackage[usenames]{color}

\begin{document}

\title{High-frequency asymptotic behavior of self-energies in quantum impurity models} 

\author{Xin Wang}
\affiliation{Condensed Matter Theory Center, Department of Physics, University of Maryland, College Park, Maryland 20742, USA}
\author{Hung The Dang}
\affiliation{Department of Physics, Columbia University, 538 West 120th Street, New York, New York 10027, USA}
\author{Andrew J. Millis}
\affiliation{Department of Physics, Columbia University, 538 West 120th Street, New York, New York 10027, USA}

\date{\today}

\begin{abstract}
We present explicit expressions for the high-frequency asymptotic behavior of electron self-energy of general quantum impurity models, which may be useful for improving the convergence of dynamical mean-field calculations and for the analytic continuation of the electron self-energy.
We also give results, expressed in more physical terms, for the two-orbital and three-orbital rotationally invariant Slater-Kanamori interactions, in order to facilitate calculations of transition metal oxides. 
\end{abstract}

\begin{comment}
71.27.+a	Strongly correlated electron systems; heavy fermions
71.10.-w	Theories and models of many-electron systems
71.10.Fd	Lattice fermion models (Hubbard model, etc.)
71.15.-m	Methods of electronic structure calculations (see also 31.15.-p Calculations and mathematical techniques in atomic and molecular physics; for electronic structure calculations of superconducting materials, see 74.20.Pq)
\end{comment}
\maketitle

Quantum Monte Carlo solutions of quantum impurity models\cite{Gull.11} play an important role in the dynamical mean-field theory (DMFT) of materials.\cite{Georges.96,Kotliar.06} In these calculations, knowledge of the high-frequency behavior of the electron self-energy is useful, both to improve the convergence of the dynamical mean-field self-consistency process\cite{Maier.05,Bluemer.PhD,Comanac.PhD} and to provide the normalization needed for analytic continuation of the electron self-energy.\cite{Wang.09} While the general procedure for deriving the high frequency behavior is known,\cite{Nolting.72,Deisz.97,Potthoff.97,Gull.11} explicit expressions seem not to have appeared except for the single-orbital Anderson impurity model\cite{Deisz.97,Potthoff.97,Freericks.09} and its cluster generalizations.\cite{Haule.07a,Haule.07b}

In this Brief Report, we present these results for general quantum impurity models. We also present the specialization to the two and three orbital models with rotationally invariant (Slater-Kanamori)\cite{Slater.36,Kanamori.63} on-site interactions, relevant to late and early transition metal oxides\cite{Werner.08,Chan.09,Wang.11,Han.11} where the physics is dominated by $e_g$ and $t_{2g}$ orbitals respectively. In supplementary material (available online) we present a Mathematica notebook which generates the needed terms and presents the resulting output. 

A quantum impurity model involves a set of ``impurity'' degrees of freedom coupled to a non-interacting bath. We 
begin by defining a general impurity model. We denote the impurity states by an index $a$ representing spin and orbital degrees of freedom and define the operator that creates an electron into impurity state $a$ as $\psi^\dagger_a$. We denote the bath states by an energy/momentum label $k$ and a spin/orbital index $\alpha$, with $c^\dagger_{k\alpha}$ the operator creating an electron into one of these state. The impurity model is the sum of a non-interacting $(H_{\rm nonint})$ and interacting $(H_{\rm int})$ terms
\begin{equation}
H=H_{\rm nonint}+H_{\rm int}
\end{equation}
with
\begin{equation}
\begin{split}
H_{\rm nonint}=&\sum_{k\alpha}\varepsilon_{k\alpha}c_{k\alpha}^\dagger c_{k\alpha}+\sum_{k\alpha b}\left(V_k^{\alpha b}c_{k\alpha}^\dagger\psi_b+H.c.\right)\\
&+\sum_{ab}E^{ab}\psi_a^\dagger\psi_b,
\end{split}
\end{equation}
and 
\begin{equation}
H_{\rm int}=\sum_{a_1a_2b_1b_2}I^{a_1a_2b_1b_2}\psi_{a_1}^\dagger\psi_{a_2}^\dagger\psi_{b_1}\psi_{b_2}.
\end{equation}

The electron self-energy $\Sigma$ is then a matrix in space of impurity model states $a$
with components $\Sigma^{ab}$. An expression for the high-frequency behavior can be obtained from examination of the short-time behavior of the electron Green's function,\cite{Potthoff.97,Comanac.PhD,Gull.11} which in turn can be expressed in terms of the commutator of the Hamiltonian with electron operators. Evaluating Eqs.~(215) and (216) of Ref.~\onlinecite{Gull.11} we find:
\begin{equation}
\lim_{z\rightarrow\infty}\Sigma^{ab}(z)=\Sigma^{ab}_\infty+\frac{\Sigma_1^{ab}}{z},
\end{equation}
with
\begin{equation}
\Sigma_\infty^{ab}=4\sum_{a_1b_1}I^{aa_1b_1b}\left\langle\psi_{a_1}^\dagger\psi_{b_1}\right\rangle,\label{series1}
\end{equation}
and
\begin{equation}
\begin{split}
\Sigma_1^{ab}=&\sum_{a_1a_2b_1b_2}K^{ab}_{a_1a_2b_1b_2}\left\langle\psi_{a_1}^\dagger\psi_{a_2}^\dagger\psi_{b_1}\psi_{b_2}\right\rangle\\
&+\sum_{a_1b_1}L^{ab}_{a_1b_1}\left\langle\psi_{a_1}^\dagger\psi_{b_1}\right\rangle-\sum_\lambda\Sigma_\infty^{a\lambda}\Sigma_\infty^{\lambda b},
\end{split}
\end{equation}
where
\begin{equation}
K^{ab}_{a_1a_2b_1b_2}=\sum_\lambda\left(4I^{a\lambda b_1b_2}I^{a_1a_2\lambda b}-16I^{aa_1b_1\lambda}I^{\lambda a_2b_2b}\right),
\end{equation}
and
\begin{equation}
L^{ab}_{a_1b_1}=-8\sum_\lambda I^{aa_1\lambda_1\lambda_2}I^{\lambda_1\lambda_2b_1b}.\label{series2}
\end{equation}
In these expressions the expectation values must in general be evaluated from the numerical solution of the quantum impurity model.

For the single-orbital Anderson impurity model,\cite{Anderson.61} the nonzero elements of the tensor $I^{a_1a_2b_1b_2}$ may be chosen as $I^{\uparrow\downarrow\downarrow\uparrow}=I^{\downarrow\uparrow\uparrow\downarrow}=U/4$ and $I^{\uparrow\downarrow\uparrow\downarrow}=I^{\downarrow\uparrow\downarrow\uparrow}=-U/4$. One can readily verify that Eqs.~\eqref{series1}-\eqref{series2} yield the well-known results that\cite{Deisz.97,Potthoff.97,Freericks.09}
\begin{align}
\Sigma_\infty^{\uparrow\uparrow}&=U\langle n_{\downarrow}\rangle,\\
\Sigma_1^{\uparrow\uparrow}&=U^2\langle n_{\downarrow}\rangle(1-\langle n_{\downarrow}\rangle),\end{align}
while $\Sigma^{\downarrow\downarrow}$ can be found by flipping all spin indices, and all ``off-diagonal'' self-energies (e.g. $\Sigma^{\uparrow\downarrow}$) are zero.

While these expressions [Eqs.~\eqref{series1}-\eqref{series2}] may be straightforwardly programmed and evaluated for general impurity models and interactions, it is useful to present results for the important special case of single-site dynamical mean-field theory of  $d$ electrons subject to rotationally invariant Slater-Kanamori interactions,\cite{Slater.36,Kanamori.63} because the results can be expressed in terms of physically relevant densities and the symmetries are automatically implemented, leading to fewer expressions to be evaluated. 

The on-site interacting part of the Slater-Kanamori Hamiltonian is traditionally expressed in terms of three parameters, $U$, $U^{'}$ and $J$, as
\begin{align}
H_{\rm int}&=\sum_\alpha Un_{\alpha\uparrow}n_{\alpha\downarrow}+\sum_{\alpha>\beta, \sigma} U'n_{\alpha\sigma}n_{\beta\overline{\sigma}}\notag\\
+&\sum_{\alpha>\beta, \sigma} (U'-J)n_{\alpha\sigma}n_{\beta\sigma}\\
-&\sum_{\alpha\neq\beta}J\left(\psi_{\alpha\downarrow}^\dagger\psi_{\beta\uparrow}^\dagger\psi_{\beta\downarrow}\psi_{\alpha\uparrow}+\psi_{\beta\uparrow}^\dagger\psi_{\beta\downarrow}^\dagger\psi_{\alpha\uparrow}\psi_{\alpha\downarrow}\right),\notag
\end{align}
where $n_{\alpha\sigma}=\psi_{\alpha\sigma}^\dagger\psi_{\alpha\sigma}$ and $\alpha$ labels the orbitals. 

The five orbital states of the $d$-manifold are degenerate in free space, but in cubic symmetry  are split into a doublet, transforming as the $e_g$ representation of the cubic group, and a triplet, transforming as the $t_{2g}$ representation. In many materials one may restrict attention to one of these two manifolds. Symmetry considerations then imply that the self-energy is diagonal in the orbital basis and, in paramagnetic states, diagonal in the spin states as well.

In the case of two-orbital model ($e_g$ system), $\alpha,\beta\in \{1,2\}$.  We can express the  high-frequency behavior for the self-energy of a spin up electron on orbital 1 in terms of expectation values of orbital densities and spin-exchange operators as
\begin{align}
\Sigma_\infty^{1\uparrow,1\uparrow}&=U\langle n_{1\downarrow}\rangle+U'\langle n_{2\downarrow}\rangle+(U'-J)\langle n_{2\uparrow}\rangle,\\
\Sigma_1^{1\uparrow,1\uparrow}&=U^2\langle n_{1\downarrow}\rangle(1-\langle n_{1\downarrow}\rangle)
+(U')^2\langle n_{2\downarrow}\rangle(1-\langle n_{2\downarrow}\rangle)\nonumber\\
&\quad+(U'-J)^2\langle n_{2\uparrow}\rangle(1-\langle n_{2\uparrow}\rangle)\nonumber\\
&\quad+2UU'(\langle n_{1\downarrow}n_{2\downarrow}\rangle-\langle n_{1\downarrow}\rangle\langle n_{2\downarrow}\rangle)\nonumber\\
&\quad+2U(U'-J)(\langle n_{1\downarrow}n_{2\uparrow}\rangle-\langle n_{1\downarrow}\rangle\langle n_{2\uparrow}\rangle)\nonumber\\
&\quad+2U'(U'-J)(\langle n_{2\uparrow}n_{2\downarrow}\rangle-\langle n_{2\uparrow}\rangle\langle n_{2\downarrow}\rangle)\nonumber\\
&\quad+J^2\left(\langle n_{1\downarrow}\rangle+\langle n_{2\downarrow}\rangle-2\langle n_{1\downarrow}n_{2\downarrow}\rangle\right)\notag\\
&\quad+2J(J-U)\left\langle \psi_{1\uparrow}^\dagger \psi_{2\downarrow}^\dagger \psi_{1\downarrow}\psi_{2\uparrow}\right\rangle\nonumber\\
&\quad+2J(U+J-2U')\left\langle \psi_{1\uparrow}^\dagger \psi_{1\downarrow}^\dagger \psi_{2\downarrow}\psi_{2\uparrow}\right\rangle,
\end{align}
and the forms for the other orbital/spin can be straightforwardly found by interchanging indices. Note that the off-diagonal elements are zero.

For the three-orbital case ($t_{2g}$ system), $\alpha,\beta$ run from $1$ to $3$. The high-frequency behavior of the self-energy for a spin up electron on orbital 1 can be similarly expressed as
\begin{widetext}
\begin{align}
\Sigma_\infty^{1\uparrow,1\uparrow}&=U\langle n_{1\downarrow}\rangle+U'(\langle n_{2\downarrow}\rangle+\langle n_{3\downarrow}\rangle)+(U'-J)(\langle n_{2\uparrow}\rangle+\langle n_{3\uparrow}\rangle),\\
\Sigma_1^{1\uparrow,1\uparrow}&=U^2\langle n_{1\downarrow}\rangle(1-\langle n_{1\downarrow}\rangle)
+(U')^2\left[(\langle n_{2\downarrow}\rangle+\langle n_{3\downarrow}\rangle)(1-\langle n_{2\downarrow}\rangle-\langle n_{3\downarrow}\rangle)+2\langle n_{2\downarrow}n_{3\downarrow}\rangle\right]\nonumber\\
&\quad+(U'-J)^2\left[(\langle n_{2\uparrow}\rangle+\langle n_{3\uparrow}\rangle)(1-\langle n_{2\uparrow}\rangle-\langle n_{3\uparrow}\rangle)+2\langle n_{2\uparrow}n_{3\uparrow}\rangle\right]+J^2(2\langle n_{1\downarrow}\rangle+\langle n_{2\downarrow}\rangle+\langle n_{3\downarrow}\rangle)\nonumber\\
&\quad+2(UU'-J^2)(\langle n_{1\downarrow}n_{2\downarrow}\rangle+\langle n_{1\downarrow}n_{3\downarrow}\rangle)-2UU'\langle n_{1\downarrow}\rangle(\langle n_{2\downarrow}\rangle+\langle n_{3\downarrow}\rangle)\nonumber\\
&\quad+2U(U'-J)(\langle n_{1\downarrow}n_{2\uparrow}\rangle+\langle n_{1\downarrow}n_{3\uparrow}\rangle-\langle n_{1\downarrow}\rangle\langle n_{2\uparrow}\rangle-\langle n_{1\downarrow}\rangle\langle n_{3\uparrow}\rangle)\nonumber\\
&\quad+2U'(U'-J)\left[\langle n_{2\uparrow}n_{2\downarrow}\rangle+\langle n_{2\downarrow}n_{3\uparrow}\rangle+\langle n_{2\uparrow}n_{3\downarrow}\rangle+\langle n_{3\uparrow}n_{3\downarrow}\rangle-(\langle n_{2\uparrow}\rangle+\langle n_{3\uparrow}\rangle)(\langle n_{2\downarrow}\rangle+\langle n_{3\downarrow}\rangle)\right]\nonumber\\
&\quad+2J(J-U)\left(\left\langle \psi_{1\uparrow}^\dagger \psi_{2\downarrow}^\dagger \psi_{1\downarrow}\psi_{2\uparrow}\right\rangle+\left\langle \psi_{1\uparrow}^\dagger \psi_{3\downarrow}^\dagger \psi_{1\downarrow}\psi_{3\uparrow}\right\rangle\right)\nonumber\\
&\quad+2J(U+J-2U')\left(\left\langle \psi_{1\uparrow}^\dagger \psi_{1\downarrow}^\dagger \psi_{2\downarrow}\psi_{2\uparrow}\right\rangle+\left\langle \psi_{1\uparrow}^\dagger \psi_{1\downarrow}^\dagger \psi_{3\downarrow}\psi_{3\uparrow}\right\rangle\right)\nonumber\\
&\quad-J^2\left(
\left\langle \psi_{2\downarrow}^\dagger \psi_{2\uparrow}^\dagger \psi_{3\downarrow}\psi_{3\uparrow}\right\rangle
+\left\langle \psi_{2\downarrow}^\dagger \psi_{3\uparrow}^\dagger \psi_{2\uparrow}\psi_{3\downarrow}\right\rangle
+\left\langle \psi_{2\uparrow}^\dagger \psi_{3\downarrow}^\dagger \psi_{2\downarrow}\psi_{3\uparrow}\right\rangle
+\left\langle \psi_{3\downarrow}^\dagger \psi_{3\uparrow}^\dagger \psi_{2\downarrow}\psi_{2\uparrow}\right\rangle
\right),
\end{align}
\end{widetext}
while the forms for other orbitals/spins can be found by permuting indices. The off-diagonal elements are also zero. In this formulation only 27 correlators (with four fermionic operators) need to be measured, rather than the $6^4=1296$ correlators arising in a straightforward implementation of the general equations above.

As the number of orbitals increases, the number of terms proliferates, and for example in a treatment of the full $5$-fold degenerate $d$ manifold more than 100 terms occur.  The expressions are too lengthy to present in the published version, but we include as supplementary information a Mathematica notebook which generates the needed terms and presents the resulting output. In the same file we have also presented the calculation for the two and three orbital cases. This code can also easily be manipulated to treat Hamiltonians with arbitrary interactions.

\section*{Acknowledgements}
We acknowledge B. Surer and N. Lin for discussions. X.W. is supported by the Condensed Matter Theory Center of University of Maryland. H.T.D. and A.J.M. are supported by NSF-DMR-1006282. Some of the results are cross-checked or calculated using the SNEG\cite{sneg} library and Wolfram Mathematica, which are also used in the supplementary information.

\end{document}